\documentclass[12pt]{article}
\usepackage{epsfig}
\usepackage{color}
\usepackage{amssymb,amsmath}
\usepackage{graphicx}
\usepackage{here}
\newcommand{\bea}{\begin{eqnarray}}
\newcommand{\eea}{\end{eqnarray}}

 \makeatletter
\def\alt{\mathrel{\mathpalette\gl@align<}}
\def\agt{\mathrel{\mathpalette\gl@align>}}
\def\gl@align#1#2{\lower.6ex\vbox{\baselineskip\z@skip\lineskip\z@
\ialign{$\m@th#1\hfil##\hfil$\crcr#2\crcr\sim\crcr}}} \makeatother

\begin{document}
\begin{flushright}
\end{flushright}
\vspace*{1.0cm}

\begin{center}
\baselineskip 20pt 
{\Large\bf 
Vacuum stability and Q-ball formation in the Type II Seesaw model
}
\vspace{1cm}

{\large 
Naoyuki Haba$^a$, \ Yasuhiro Shimizu$^a$ \ and \ Toshifumi Yamada$^b$
} \vspace{.5cm}

{\baselineskip 20pt \it
$^a$Department of Physics, Osaka Metropolitan University,
Osaka 558-8585, Japan
\\
$^b$Department of Physics, Yokohama National University,
Yokohama 240-8501,Japan
}

\vspace{.5cm}

\vspace{1.5cm} {\bf Abstract} \end{center}
We investigate vacuum stability and Q-ball formation in the Type II seesaw model by considering the effective potential for scalar fields, taking into account renormalization effects. We find that the quartic coupling for the triplet Higgs can vanish at a high energy scale, creating a flat direction where Q-ball formation can occur. If Q-balls are produced, they eventually decay into leptons via neutrino Yukawa couplings with the triplet Higgs. If this decay occurs above the electroweak scale, the leptogenesis scenario can work, and the baryon number is produced via the sphaleron effect. We show that there are parameter regions where the above scenario occurs, taking into account phenomenological constraints.
\thispagestyle{empty}

%\bigskip
\newpage

%\addtocounter{page}{-1}
\setcounter{footnote}{0}
%%%%%%%%%%%%%%%%%%%%%%%%%%
%\baselineskip 36pt
% Main body
%%%%%%%%%%%%%%%%%%%%%%%%%%
\baselineskip 18pt
%%%%%%%%%%%%%%%%%%%%%%%%%%
%

\section{Introduction}

The Standard Model (SM) of particle physics has been a successful theory for over four decades. However, it is still incomplete and some of the most crucial questions in particle physics today, 
such as the origin of neutrino masses and the nature of dark matter, are not addressed by the SM. 
The Seesaw mechanism is one of the most widely studied extensions of the SM that can address the neutrino mass problem. The mechanism proposes the addition of a heavy particle spectrum, which can explain the smallness of the observed neutrino masses via a suppression factor proportional to the inverse of their mass scale. There are three types of the Seesaw mode, I \cite{Yanagida:1979as,Gell-Mann:1979vob,Mohapatra:1979ia}, II \cite{Magg:1980ut, Cheng:1980qt,Lazarides:1980nt,Mohapatra:1980yp}, III \cite{Foot:1988aq}.

The type II Seesaw model is a particular extension of the Seesaw mechanism that introduces a scalar triplet. This model can explain the smallness of neutrino masses through a combination of the Seesaw mechanism and the triplet vacuum expectation value. The scalar triplet can interact with the SM Higgs boson and potentially affect its stability. In Ref. \cite{Haba:2016zbu},
they studied the vacuum stability of the SM Higgs potential in the type II Seesaw model and found that a large coupling between the SM Higgs and the triplet Higgs is required to ensure vacuum stability.
In addition, the neutrino sector and baryon asymmetry of the universe may be related via leptogenesis 
in this model.  However, it was shown that the thermal leptogenesis scenario does not work in the minimal Type II Seesaw model \cite{Ma:1998dx,Hambye:2005tk}.

One possible solution to this problem is the formation of Q-balls \cite{Coleman:1985ki}, which are non-topological solitons that can arise in scalar field theories with a conserved global charge.  In the context of the Seesaw mechanism, Q-balls can form from the coherent oscillations of the triplet Higgs, which carries the lepton number. If Q-balls are produced, they eventually decay into leptons via neutrino Yukawa couplings with the triplet Higgs. If this decay occurs above the electroweak scale, the leptogenesis scenario can work, and the baryon number is produced via the sphaleron effect.
There has been considerable interest in the formation of Q-balls in the context of the type II Seesaw model. To achieve the Q-balls formation, the scalar potential should have a flat direction. In non-supersymmetric models, it is not easy to have a flat direction in scalar potentials. 
In Ref. \cite{Barrie:2022cub}, they examined the type II Seesaw model with non-minimal couplings to gravity, and the flat direction is obtained since the triplet Higgs  is assumed as a component of inflation, whose potential must be very flat to achieve the inflation.

In this paper, we consider the vacuum stability and Q-ball formations in the type II Seesaw model with a scalar triplet taking account of the renormalization effects. If the neutrino Yukawa coupling is 
sufficiently large, the self-coupling of the triplet Higgs can vanish at a high energy scale
due to the renormalization effects and the flat direction appears along to  the triplet Higgs directions. We consider the effective potential improved by renormalization group equations (RGEs) at the one-loop level and investigate the vacuum stability. We find parameter regions where the self-coupling of the triplet Higgs  vanishes taking account of various phenomenological constraints.

The rest of the paper is organized as follows. In Section II, we describe the type II Seesaw model and its scalar sector. In Section III, we discuss the effective potential and vacuum stability. In Section IV, we show the numerical results. Finally, we summarize our conclusions in Section VI.

\section{Type II Seesaw Model}
We consider the minimal type-II seesaw model where an $SU(2)_L$ triplet scalar field $\Delta$ with hypercharge 1 is introduced in addition to the SM fields.
The triplet field $\Delta$ can be parametrized as follows,
\begin{eqnarray}
	\Delta = \frac{\sigma^i}{\sqrt 2}\Delta_i 
		= \left( \begin{array}{cc}
			\delta^+/\sqrt 2 & \delta^{++}\\
			\delta^0 & -\delta^+/\sqrt 2
		\end{array} \right).
\end{eqnarray}
Here $\Delta_1=(\delta^{++}+\delta^0)/\sqrt 2,~\Delta_2=i(\delta^{++}-\delta^0)/\sqrt 2,~\Delta_3=\delta^+$.  The Yukawa interactions for this model are given by 
\begin{eqnarray}
	{\cal L}_Y &=& {\cal L}_Y^{\rm SM}
	- \frac{1}{\sqrt 2}\left(Y_\Delta\right)_{ij} L_i^{\sf T}Ci\sigma_2\Delta L_j+{\rm H.c.}\, ,
 \label{YNU}       
\end{eqnarray}
where $L_i$ are the left-handed lepton doublets and $C$ is the Dirac charge conjugation matrix with respect to the Lorentz group.
The scalar potential  can be written by
\begin{eqnarray}
	{\cal V}(\Phi, \Delta) &=& - m_\Phi^2 \Phi^\dagger \Phi + \frac{\lambda}{2} (\Phi^\dagger \Phi)^2
		+ M_\Delta^2 {\rm Tr}(\Delta^\dagger \Delta)
		+ \frac{\lambda_1}{2} \left[ {\rm Tr}(\Delta^\dagger \Delta) \right]^2 \nonumber \\
	&& + \frac{\lambda_2}{2} \left( \left[ {\rm Tr}(\Delta^\dagger \Delta) \right]^2
			- {\rm Tr} \left[ (\Delta^\dagger \Delta)^2 \right] \right)
		+ \lambda_4 (\Phi^\dagger \Phi) {\rm Tr}(\Delta^\dagger \Delta)
		+ \lambda_5 \Phi^\dagger [\Delta^\dagger, \Delta] \Phi \nonumber \\
	&& + \left( \frac{\Lambda_6}{\sqrt{2}} \Phi^{\rm T} i \sigma_2 \Delta^\dagger \Phi + {\rm H.c}. \right),
\label{potential}
\end{eqnarray}
where $\Phi$ is the SM Higgs doublet.
The triplet Higgs can carry the lepton number of $-2$ and  this potential contains a rich structure that allows for a variety of phenomena, such as electroweak symmetry breaking, neutrino mass generation, and the formation of Q-balls.

We consider the parameter regions with $m_\Phi^2>0$ to achieve the electroweak symmetry breaking. At the potential minimum, the masses of $\Phi$ and $\Delta$ are given by
\begin{eqnarray}
	m_\Phi^2 &=& \frac{1}{2}\lambda v^2 - \Lambda_6 v_\Delta
		+ \frac{1}{2} (\lambda_4-\lambda_5) v_\Delta^2, \\
	M_\Delta^2 &=& \frac{1}{2} \frac{\Lambda_6 v^2}{v_\Delta}
		- \frac{1}{2} (\lambda_4-\lambda_5) v^2 - \frac{1}{2} \lambda_1 v_\Delta^2,
\label{M_Delta}
\end{eqnarray}
 where $v=246$ GeV is the VEV of the SM Higgs and $v_\Delta$ is the VEV of neutral components of  $\Delta$. In the limit $v_\Delta \ll v$, $v_\Delta$ can be written by 
\begin{eqnarray}
	v_\Delta \approx \frac{\lambda_6 M_\Delta v^2}{2 M_\Delta^2 + v^2(\lambda_4- \lambda_5)},
\end{eqnarray}
where we define $\lambda_6 \equiv \Lambda_6 /M_\Delta$.
From the Eq.(\ref{YNU}) the neutrino mass matrix is given by
\begin{eqnarray}
	(M_\nu)_{ij} = v_\Delta (Y_\Delta)_{ij}
		\approx \frac{\lambda_6 M_\Delta v^2}{2 M_\Delta^2 + v^2(\lambda_4- \lambda_5)} (Y_\Delta)_{ij},
\end{eqnarray}

 The magnitude of the triplet VEV is constrained by both theoretical and experimental considerations. The theoretical lower bound on the triplet VEV is  $\mathcal{O}(10^{-2})$ eV, 
 since the Yukawa coupling should be less than $\mathcal{O}(1)$ from the perturbativity bound.
 On the other hand, the upper bound on the triplet VEV is obtained by the $\rho$  parameter, the current experimental data implies that $v_\Delta\lesssim 5$ GeV \cite{Haba:2016zbu}.
 Therefore the allowed range is  roughly given by
\begin{equation}
 10^{-2}~ \mathrm{eV} \lesssim v_\Delta \lesssim 5 ~ \mathrm{GeV}.
\end{equation}

\section{Vacuum stability and Q-ball formation}
Let us investigate the vacuum stability in the Type II Seesaw model.  
To study the vacuum stability of the scalar potential, we consider the effective potential improved by RGEs at the one-loop level \cite{Sher:1988mj,Branco:2011iw}, which is written by
\begin{eqnarray}
	{\cal V}_{(\mathrm{eff})}(\Phi, \Delta) &=& - m_\Phi^{'2}(t) \Phi^\dagger \Phi + \frac{\lambda'(t)}{2} (\Phi^\dagger \Phi)^2
		+ M_\Delta^{'2}(t) {\rm Tr}(\Delta^\dagger \Delta)
		+ \frac{\lambda'_1(t)}{2} \left[ {\rm Tr}(\Delta^\dagger \Delta) \right]^2 \nonumber \\
	&& + \frac{\lambda'_2(t)}{2} \left( \left[ {\rm Tr}(\Delta^\dagger \Delta) \right]^2
			- {\rm Tr} \left[ (\Delta^\dagger \Delta)^2 \right] \right)
		+ \lambda'_4(t) (\Phi^\dagger \Phi) {\rm Tr}(\Delta^\dagger \Delta)
		+ \lambda'_5(t) \Phi^\dagger [\Delta^\dagger, \Delta] \Phi \nonumber \\
	&& + \left( \frac{\Lambda'_6(t)}{\sqrt{2}} \Phi^{\rm T} i \sigma_2 \Delta^\dagger \Phi + {\rm H.c}. \right),
\label{potential2}
\end{eqnarray}
where the effective masses and couplings are given by
\begin{eqnarray}
 && m_\Phi^{'2}(t) = m_\Phi^{2} G_\Phi^{2}(t),~~\lambda'(t) = \lambda G_\Phi^{4}(t)
  \nonumber \\
 && M_\Delta^{'2}(t)=M_\Delta^{2} G_\Delta^{2}(t),~~\lambda'_1(t) = \lambda_1 G_\Delta^{4}(t),~~\lambda'_2(t) = \lambda_2 G_\Delta^{4}(t),
  \nonumber \\
&& \lambda'_4(t) = \lambda_4 G_\Phi^{2}(t)G_\Delta^{2}(t),~~\lambda'_5(t) = \lambda_5 G_\Phi^{2}(t)G_\Delta^{2}(t),~~\Lambda'_6(t) = \Lambda_6 G_\Phi^{2}(t)G_\Delta(t).
\end{eqnarray}
Here $t=\log(\Phi/M)$ with a renormalization scale $M$ and
\begin{eqnarray}
G_i(t)=\exp\left[-\int_0^tdt'\gamma_i(t')\right],~~~(i=\Phi,\Delta),
\end{eqnarray}
where $\gamma_\Phi$ and $\gamma_\Delta$ are the anomalous dimensions for $\Phi$ and $\Delta$, respectively. The one-loop beta functions of coupling constants are given in \cite{Haba:2016zbu}

\begin{eqnarray}
16\pi^2 \beta_\lambda &=&
	\lambda \left[ 12\lambda - \left( \frac{9}{5}g_1^2 + 9g_2^2 \right) + 12y_t^2 \right]
	+ \frac{9}{4} \left( \frac{3}{25}g_1^4 + \frac{2}{5}g_1^2g_2^2 + g_2^4 \right) +6\lambda_4^2 + 4\lambda_5^2 - 12y_t^4,\nonumber \\
%	& + &6\lambda_4^2 + 4\lambda_5^2 - 12y_t^4,
 \nonumber\\
    16\pi^2 \beta_{\lambda_1} &=&
	\lambda_1 \left[ 14\lambda_1 + 4 \lambda_2
	- \left( \frac{36}{5} g_1^2 + 24g_2^2 \right)
	+ 4 {\rm tr} \left[{\bf S}_\Delta \right] \right]
	+ \frac{108}{25}g_1^4 + \frac{72}{5}g_1^2g_2^2 + 18g_2^4 \nonumber\\
	& +& 2 \lambda_2^2 + 4 \lambda_4^2 + 4\lambda_5^2
	- 8 {\rm tr} \left[{\bf S}_\Delta^2 \right], \nonumber\\
16\pi^2 \beta_{\lambda_2} &=&
	\lambda_2 \left[ 12 \lambda_1 + 3 \lambda_2 
	- \left( \frac{36}{5}g_1^2 + 24g_2^2 \right)
	+ 4 {\rm tr} \left[{\bf S}_\Delta  \right] \right]
	- \frac{144}{5}g_1^2g_2^2 + 12g_2^4 -8 \lambda_5^2 + 8 {\rm tr} \left[{\bf S}_\Delta^2 \right],\nonumber\\
%	& -	& 8 \lambda_5^2 + 8 {\rm tr} \left[{\bf S}_\Delta^2 \right],  \nonumber\\
%
16\pi^2\ \beta_{\lambda_4} &=&
	\lambda_4 \left[ 6 \lambda + 8 \lambda_1 + 2 \lambda_2 + 4\lambda_4
	- \left( \frac{9}{2}g_1^2 + \frac{33}{2}g_2^2 \right) 
	+ 6 y_t^2 + 2 {\rm tr} \left[{\bf S}_\Delta \right] \right] \nonumber\\
	& +	& \frac{27}{25}g_1^4 + 6g_2^4
	+ 8 \lambda_5^2 - 4 {\rm tr}\left[ {\bf S}_\Delta^2 \right], \nonumber\\
16\pi^2 \beta_{\lambda_5} &=&
	\lambda_5 \left[ 2 \lambda + 2\lambda_1 - 2\lambda_2 + 8 \lambda_4
	- \left( \frac{9}{2}g_1^2 + \frac{33}{2}g_2^2 \right)
	+ 6 y_t^2 + 2 {\rm tr}\left[{\bf S}_\Delta \right] \right]
	- \frac{18}{5}g_1^2g_2^2 +4 {\rm tr}\left[{\bf S}_\Delta^2 \right],\nonumber\\
%	& +	& 4 {\rm tr}\left[{\bf S}_\Delta^2 \right],
 \nonumber\\
 16\pi^2 \beta_{{\bf S}_\Delta} 	&=	&
	{\bf S}_\Delta \left[ 6\, {\bf S}_\Delta - 3 \left( \frac{3}{5} g_1^2 + 3 g_2^2 \right)
	+ 2 {\rm tr}[{\bf S}_\Delta] \right],
 \label{beta}
\end{eqnarray}
where we define ${\bf S}_\Delta=Y_\Delta^\dagger Y_\Delta $. Using these RGEs, we can  determine whether any of the couplings become negative, indicating vacuum instability. In the case of the Type II Seesaw model, it is known that the quartic coupling $\lambda_\Delta$ can become negative at high energy scales, leading to potential instability. Therefore, it is important to study the vacuum stability of the scalar potential in this model, in order to determine the allowed range of parameter values that lead to a stable vacuum.

It is known that the current experimental values of the SM parameters suggest that the Higgs vacuum is only metastable in the SM, which can be seen in the running of the renormalization group equations of the Higgs quartic coupling in the SM. In the Type II Seesaw model, the necessary and sufficient conditions for the vacuum stability is givey by \cite{Arhrib:2011vc}
\begin{eqnarray}
        &&  \lambda \geq 0,\quad \lambda_1 \geq 0,\quad 2 \lambda_1 + \lambda_2 \geq 0, \nonumber \\
        &&  \lambda_4 + \lambda_5 + \sqrt{\lambda \lambda_1} \geq 0,\quad
        \lambda_4 + \lambda_5 + \sqrt{\lambda \left( \lambda_1 + \frac{\lambda_2}{2} \right)} \geq 0, \nonumber \\
        &&  \lambda_4 - \lambda_5 + \sqrt{\lambda \lambda_1} \geq 0,\quad
        \lambda_4 - \lambda_5 + \sqrt{\lambda \left( \lambda_1 + \frac{\lambda_2}{2} \right)} \geq 0. 
\label{stability}
\end{eqnarray}
It was shown that a sufficient coupling between the SM Higgs and the triplet Higgs  ensures vacuum stability up to the Planck scale \cite{Haba:2016zbu}. We focus on the parameter regions where $\lambda_1$ vanishes at a high energy scale and the flat direction appears along the triplet Higgs.

\section{Numerical Results}
We consider the vacuum stability by solving the RGEs numerically. For the input values of the SM parameters, we use the following experimental data \cite{ParticleDataGroup:2022pth}.
\begin{eqnarray}
       m_t&=&172.76\pm 0.30~\mathrm{GeV}.
 \\      
       m_h&=&125.25\pm 0.17~\mathrm{GeV}.
 \\    
    \alpha_s(M_Z)&=& 0.1179\pm0.0009.
 \label{input}
\end{eqnarray}
With the SM model inputs, we solve the 1-loop RGEs in the SM up to the scale $M_\Delta$ and 
solve the 1-loop RGEs in the Type II seesaw model above this scale. For the neutrino parameters,
we use the experimental data \cite{Esteban:2020cvm},
\begin{eqnarray}
       \sin^2\theta_{12}&=&0.304^{+0.013}_{-0.012},~~~~~\Delta m^{21}=7.42^{+0.21}_{-0.20}\times 10^{-5}\mathrm{eV}^2,
 \\      
       \sin^2\theta_{13}&=&0.570^{+0.018}_{-0.024},~~~~~\Delta m^{32}=2.514^{+0.028}_{-0.027}\times 10^{-3}\mathrm{eV}^2,
 \\    
       \sin^2\theta_{13}&=&0.02221^{+0.000068}_{-0.00062},~~~\delta_{CP}=195^{\circ +51}_{-25},
 \label{input2}
\end{eqnarray}
for normal ordering of nutrino masses and $m_{\nu_1}=0$.

\begin{figure}[htbp]
\begin{center}
\includegraphics[width=15cm]{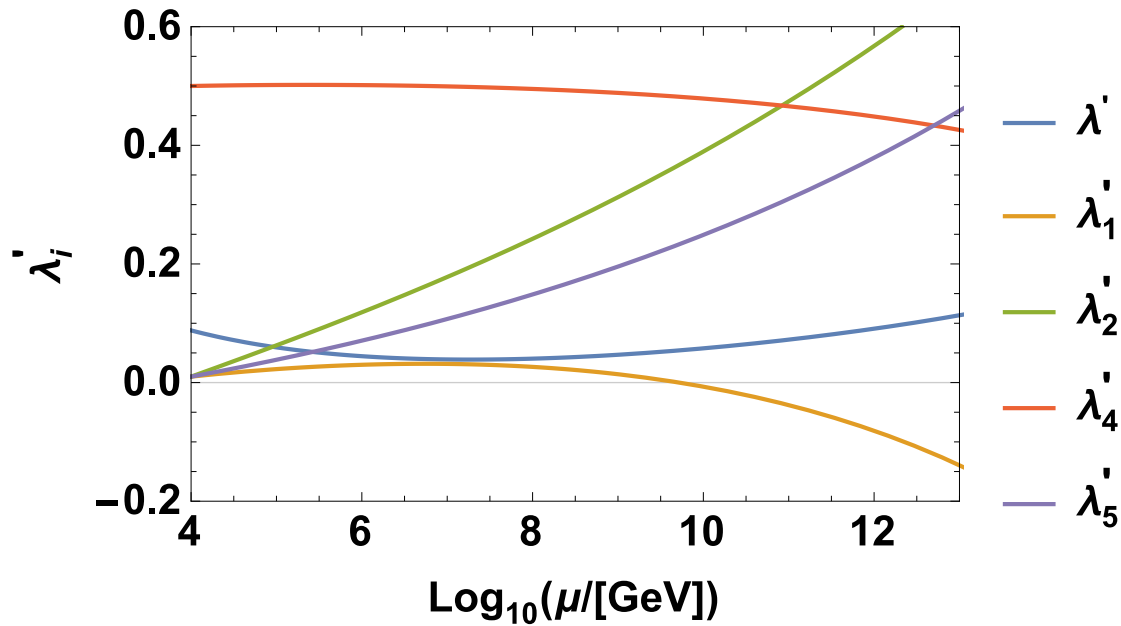}
\caption{The scale dependence of effective quartic couplings. We take $\lambda_4$ = 0.45 and $\lambda_1$ = $\lambda_2$ = $\lambda_5$ = 0.01 at $\mu=M_\Delta=10^4$ GeV and $v_\Delta=0.059$ eV.}
\label{figRGE}
\end{center}
\end{figure}

We have searched the parameter regions where $\lambda_1$ vanishes at a high energy scale while
other quartic couplings are positive to ensure the vacuum stability in Eqs.(\ref{stability}). 
To vanish $\lambda_1$ at a high energy scale, the beta function should be negative.
From the RGEs in Eqs.(\ref{beta}), the beta function for $\lambda_1$ can be negative for 
sufficiently large neutrino Yukawa couplings. However, the large neutrino Yukawa coupling
also gives a negative contribution to the beta function of $\lambda_4$. In addition, to avoid
the negative $\lambda$ for the SM Higgs, a sufficiently large $\lambda_4$ is required.
Unfortunately, we cannot find any parameter regions where $\lambda_1$ vanishes at a high energy scale while other quartic couplings are positive. 
However, we find the parameter regions where $\lambda_1$  vanishes at a high energy scale and eventually becomes negative at the higher scale while other quartic couplings are positive up to the Planck scale. As an example, the scale dependence of the effective couplings is shown in Fig.\ref{figRGE}. 
Here we take input values $\lambda_1=\lambda_2=\lambda_5=0.01$, $\lambda_4=0.45$ at $\mu=M_\Delta=10$~TeV, and $v_\Delta=0.059$ eV. We can see that $\lambda_1'$ decreases monotonically and  becomes negative at $\mu\simeq 10^{10}$ GeV. On the other hand, $\lambda'$ has a local minimum around $\mu\simeq 10^8$ GeV. This arises because the top Yukawa coupling only affects the beta function of $\lambda$ and not $\lambda_1$.

Since $\lambda_1'$ vanishes at a scale $\mu\sim 10^{10}$ GeV, the flat direction for the triplet Higgs appears around this scale. However, the $\lambda_1'$ becomes negative above this scale, which leads to an unbounded potential.   To stabilize the potential, we assume that there are higher dimensional operators, such as $(c_n/M) [\mathrm{Tr}(\Delta^\dagger\Delta)]^n$ with $c_n>0$. The flat direction for the triplet Higgs has the lepton numbers and the Q-ball with lepton numbers can form. The detailed calculation of Q-ball formation is beyond the scope of this paper.  If  Q-balls are formed, they can decay by the triplet Yukawa coupling, and lepton asymmetry is generated in the SM sector. If this scale is above the electroweak scale, the leptogenesis occurs by the sphaleron effect. 

It is known that there are constraints from  washout processes in the Type II seesaw model\cite{Barrie:2022cub}. At the beginning of the radiation epoch, the triplet Higgs rapidly thermalizes due to a large reheating temperature in this scenario, which requires considering possible washout processes. The first requirement is that the process $LL\leftrightarrow HH$ is never in thermal equilibrium, and the limit on the triplet Higgs mass is 
\begin{equation}
    m_\Delta <10^{12} ~\mathrm{GeV}~~ \mathrm{for}~~ m_\nu=0.05 \mathrm{eV}.
\end{equation}
The other dangerous processes are $LL\leftrightarrow\Delta$ and 
$HH\leftrightarrow\Delta$, which must not co-exist, In order to realize this condition 
\begin{equation}
    v_\Delta \lesssim 10^{-5}{\rm GeV} \left( \frac{m_\Delta}{1{\rm TeV}}\right)^{-1/2}
\end{equation}

The lepton flavor violating (LFV) decays can be generated in the Type II Seesaw model and can be constrained by various LFV experiments, such as $\mu \to e\gamma$, $\mu \to 3 e$ decays. In particular, the  $\mu \to 3 e$ decay is generated by the doubly-charged Higgs boson $H^{++}$ diagrams at the tree level and the branching ratio is calculated as \cite{Dinh:2012bp}
\begin{equation}
B (\mu^+ \to e^+ e^- e^+) = \frac{|(Y_\Delta)_{\mu e} (Y_\Delta)^\dagger_{ee}|^2}{4 G_F^2 M_{H^{++}}^4}\,~,
\end{equation}  
where the $H^{++}$ mass is given by
\begin{equation}
M_{H^{++}}=M_\Delta^2+\frac{1}{2}(\lambda_4+\lambda_5)v^2
		+\frac{1}{2}(\lambda_1+\lambda_2)v_\Delta^2.
\end{equation}  
The current experimental upper bound is obtained as \cite{SINDRUM:1987nra}
\begin{equation}
 B(\mu^+ \to e^+ e^- e^+) < 1.0 \times 10^{-12},
\end{equation}
and the bound is expected to reach $5.2\times 10^{-15}$ at the future experiment \cite{Perrevoort:2018ttp}.
Although the LFV decays depend on the various neutrino parameters \cite{Dinh:2012bp}\cite{Barrie:2022ake}, conducting a detailed analysis is beyond the scope of this paper. For simplicity, we consider the central values of the experimental neutrino data and assume vanishing Majorana phases. Under these assumptions, we find that the $\mu\to 3e$ search provides the most stringent bound among various current LFV experiments.

 The triplet Higgs mass is directly constrained by collider experiments. The doubly-charged Higgs boson search at the LHC gives $M_\Delta \gtrsim 800$ GeV \cite{ATLAS:2017xqs}\cite{Cai:2017mow} and
 the future 100 TeV collider, with an integrated luminosity of 50 ab$^{-1}$, is expected to achieve a 5$\sigma$ sensitivity to $M_\Delta=7$ TeV \cite{ Du:2018eaw}\cite{Fuks:2019clu}.

\begin{figure}[htbp]
\begin{center}
  \includegraphics[width=15cm]{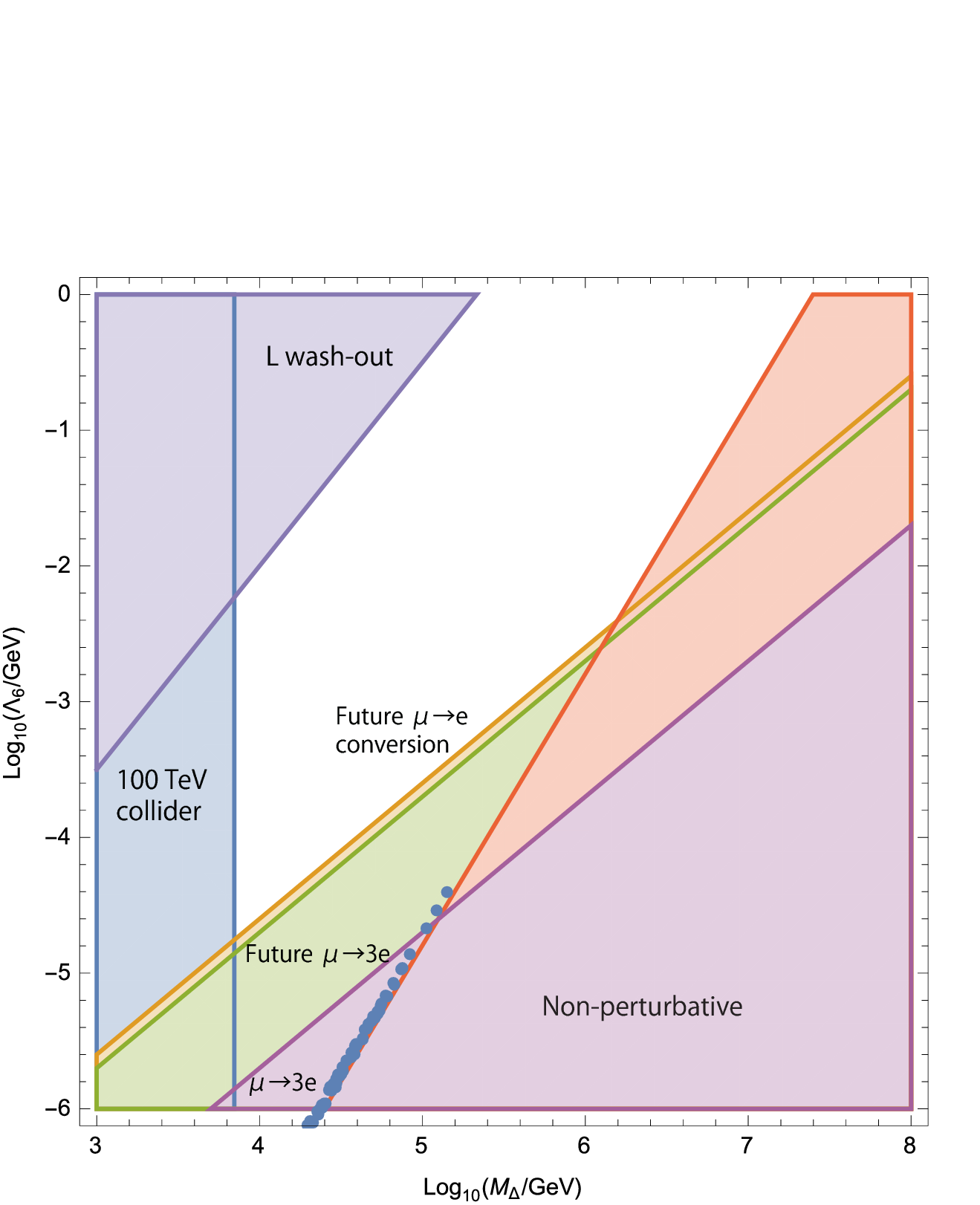}
\caption{Scatter plots on the $M_\Delta$-$\Lambda_6$ plane,
in which the $\lambda_1'$ vanishes or becomes negative at a high energy scale and  the vacuum stability conditions are satisfied. We take the input parameters as $\lambda_1'=\lambda_2'=\lambda_6'=0.01$ and $\lambda_1'=0.45$ at $\mu=M_\Delta=10^4$ GeV. We show the excluded regions by non-perturbativity of the triplet Yukawa couplings, the $\mu\to 3e$ experiments, the $\mu\to e$ conversion experiments, the lepton number washout, and the direct search at the 100 TeV collider, $M_\Delta>7$ TeV.}
\label{figcontour}
\end{center}
\end{figure}

In Fig.\ref{figcontour}, we show the scatter plot on the $M_\Delta$-$\Lambda_6$ plane,
in which the $\lambda_1'$ vanishes or becomes negative at a high energy scale and  the vacuum stability conditions are satisfied. We set the input parameters to $\lambda_1'=\lambda_2'=\lambda_6'=0.01$, with $\lambda_1'=0.45$ at $\mu=M_\Delta=10^4$ GeV.  We also impose the perturbativity conditions that all quartic couplings are less than
$4\pi$ up to the Planck scale. This figure shows that there are points in small regions near the non-perturbative Yukawa coupling region and most of the parameter regions have been already excluded by the current $\mu\to 3e$ experiment.
 Furthermore, the future $\mu\to 3e$ experiment holds the potential to explore the remaining points. Additionally, the future $\mu\to e$ conversion experiment at COMET, with an experimental sensitivity reaching $2.6 \times 10^{-17}$ \cite{COMET:2018wbw}, can also investigate these remaining points.

\section{Summary}

We have investigated vacuum stability and Q-ball formation in the Type II seesaw model by analyzing the RGE-improved effective potential at the 1-loop level. We found that the effective quartic coupling of the triplet Higgs boson, $\lambda_1'$, can vanish if the neutrino Yukawa coupling is sufficiently large, due to the negative contribution of the Yukawa coupling to the beta function of $\lambda_1'$. Our analysis revealed the existence of parameter regions where $\lambda_1'$ can vanish at high energy scales, while other quartic couplings remain positive. However, above a certain energy scale, $\lambda_1'$ becomes negative, causing the potential to be unbounded below along the triplet Higgs direction. To avoid this problem, we considered the possibility of introducing higher-dimensional interactions, which can bound the potential from below. In this case, the potential becomes very flat along the triplet Higgs direction at $\lambda_1'\simeq 0$, allowing for the formation of Q-balls with a lepton charge. If Q-balls are produced, they eventually decay into leptons via neutrino Yukawa couplings with the triplet Higgs. This process can lead to the leptogenesis scenario if it occurs above the electroweak scale.
We showed that there are parameter regions where the above scenario can occur, taking into account phenomenological constraints. Our analysis suggests that the allowed parameter regions are within the reach of the future $\mu\to 3e$ and $\mu\to e$ conversion experiments.

\section*{Acknowledgement}

This work is partially supported by Scientific Grants by the Ministry of Education, Culture, Sports, Science and Technology of Japan, Nos. 17K05415, 18H04590 and 19H051061 (NH), and No. 19K147101 (TY).

\end{document}